\documentclass[reprint, superscriptaddress, secnumarabic, amssymb, nobibnotes, aps, prl]{revtex4-1}

\usepackage{graphicx}
\usepackage{epstopdf}
\usepackage[T1]{fontenc}
\usepackage[latin9]{inputenc}
\usepackage{amsbsy}
\usepackage{gensymb}
\usepackage[T1]{fontenc}
\usepackage[latin9]{inputenc}
\usepackage{amsmath}
\usepackage{amssymb}
\usepackage{bbm}
\usepackage{braket}
\usepackage{xcolor}
\allowdisplaybreaks
\usepackage{graphicx}
\usepackage[colorlinks=true]{hyperref}  
\hypersetup{
    bookmarks=true,         
    unicode=false,          
    pdftoolbar=true,        
    pdfmenubar=true,        
    pdffitwindow=false,     
    pdfstartview={FitH},    
    pdftitle={x},    
    pdfauthor={SM},     
    pdfsubject={},   
    pdfcreator={},   
    pdfproducer={}, 
    pdfkeywords={} {} {}, 
    pdfnewwindow=true,      
    colorlinks=true,       
    linkcolor=blue, 
    citecolor=blue,        
    filecolor=magenta,      
    urlcolor=blue           
} 
\usepackage[normalem]{ulem}

\renewcommand{\approx}{\simeq}

\begin{document}
\title{\textrm{Room Temperature Ferrimagnetism, Magnetodielectric and Exchange Bias Effect in CoFeRhO$_4$}}
\author{P. Mohanty}
\affiliation{Indian Institute of Science Education and Research Bhopal, Bhopal, 462066, India}
\author{N. Sharma}
\affiliation{School of Physics and Materials Science, Thapar Institute of Engineering and Technology, Patiala 147004, India}
\author{D. Singh}
\affiliation{Indian Institute of Science Education and Research Bhopal, Bhopal, 462066, India}
\author{Y. Breard}
\affiliation{Laboratoire CRISMAT, UMR 6508 CNRS ENSICAEN, 6 bd du Marechal Juin, 14050 Caen Cedex 4, France}
\author{D. Pelloquin}
\affiliation{Laboratoire CRISMAT, UMR 6508 CNRS ENSICAEN, 6 bd du Marechal Juin, 14050 Caen Cedex 4, France}
\author{S. Marik}
\email[]{soumarik@thapar.edu}
\affiliation{School of Physics and Materials Science, Thapar Institute of Engineering and Technology, Patiala 147004, India}
\author{R. P. Singh}
\email[]{rpsingh@iiserb.ac.in}
\affiliation{Indian Institute of Science Education and Research Bhopal, Bhopal, 462066, India}

\begin{abstract}
\begin{flushleft}

\end{flushleft}
Geometrically frustrated structures combined with competing exchange interactions that have different magnitudes are known ingredients for achieving exotic properties. Herein, we studied detailed structural, magnetic, thermal (specific heat), magneto-dielectric, and magnetic exchange bias properties of a mixed 3d - 4d spinel oxide with composition CoFeRhO$_4$. Detailed magnetization, heat capacity, and neutron powder diffraction studies (NPD) highlight long-range ferrimagnetic ordering with an onset at 355 K. The magnetic structure is established using a ferrimagnetic model (collinear-type) that has a propagation vector k = 0, 0, 0. The magneto-dielectric effect appears below the magnetic ordering temperature, and the exchange bias (EB) effect is observed in field cooled (FC) conditions below 355 K. The magneto-dielectric coupling in CoFeRhO$_4$ originates due to the frustration in the structure, collinear ferrimagnetic ordering, and uncompensated magnetic moments. The unidirectional anisotropy resulting from the uncompensated magnetic moments causes the room-temperature exchange bias effect. Remarkably, the appearance of technologically important properties (ferromagnetism, magnetodielectric effect, and EB) at room temperature in CoFeRhO$_4$ indicates its potential use in sensors or spintronics.
\end{abstract}
\maketitle
\section{Introduction}
\begin{figure}
\includegraphics[width=0.90\columnwidth]{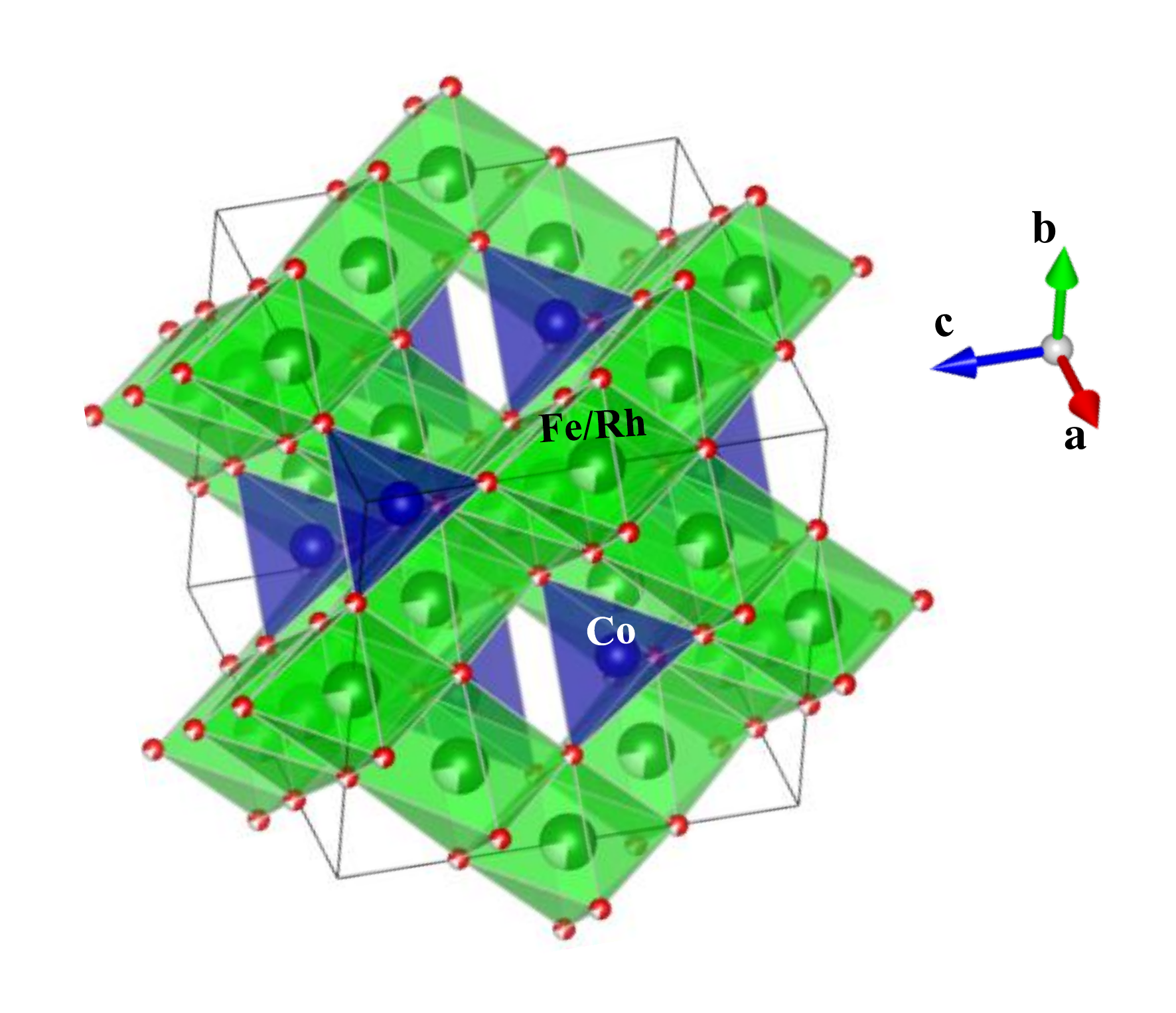}
\caption{\label{Figure 1:SRT}crystal structure for the CoFeRhO$_4$. Green spheres represent the Fe/Rh cations (Octahedral B sites in the AB$_2$O$_4$ structure), Blue spheres represent the Co cations (Tetrahedral A sites), and red spheres represent oxygen.}
\end{figure}

Spinel oxides are an intriguing class of materials, not only for their potentiality for a wide range of applications but also because of a variety of new and exciting physics (such as frustrated magnetism, multiferroic properties, orbital glass system, spintronics applications, and spin-orbital liquids) that continues to arise from the strong interactions among spin, orbital, and structural degrees of freedom \cite{1,2,3,4,5,6,7,8,9,10,11,12,13,14}. They have a unique structure with a general formula of AB$_2$O$_4$, where A and B are metal ions (Fig.\ref{Figure 1:SRT}). This structure comprises an array of metal cations in octahedral and tetrahedral coordination, surrounded by oxygens, creating two sets of magnetic sublattices \cite{15}.  B cations generally form a pyrochlore-like lattice by residing in the octahedral sites and originate frustrated magnetic interactions \cite{16, 17, 18, 19}. On the other hand, the metal A ions occupy the tetrahedral sites (eightfold) and construct a diamond lattice \cite{20,21,22,23,24}. This bipartite lattice can be interpreted as two face-centered interpenetrated cubic (fcc) sublattices. These sublattices are shifted diagonally by one-quarter. Variation of magnetic and nonmagnetic cations on the tetrahedral A sites and the octahedral B sites can originate complex magnetic interactions by affecting the magnitudes of superexchange interactions (J$_{AA}$, J$_{BB}$, and J$_{AB}$). Therefore, exotic properties (and states) such as a spiral spin liquid phase \cite{25}, unique, glassy magnetic behavior \cite{21, 26}, and spin-orbital liquids \cite{27, 28} emerge in spinel materials. The pyrochlore lattice is a fertile playground for theoretical and experimental research to explore new physics.  At the same time, the bipartite diamond lattice (in spinel systems) is a fruitful platform for realizing exotic quantum behavior. In particular, the bipartite nature of the diamond lattice can be useful in designing the recently proposed 3D topological paramagnetism for the frustrated S = 1 diamond lattice \cite{21,22}.

In addition to geometrical frustration and competing exchange interactions having different magnitudes, spin-orbit coupling (SOC) is a known ingredient that favors more exotic spin order and dynamics. In this connection, 4d and 5d elements containing oxides attracted remarkable research interest in recent times. Being spatially more extended (4d/5d orbitals), the on-site Coulomb repulsion energy (U) is smaller in 4d and 5d elements than their 3d analogues. However, in the spinel oxide family, only one 5d-containing material has been reported so far. The iridium-containing spinel oxide with composition Cu[Ir$_{1.5}$Cu$_{0.5}$]O$_4$ highlights a highly frustrated magnetic state \cite{29}. Among 4d-rhodium-based diamond lattice spinel oxides, in diamond-lattice Heisenberg antiferromagnet CoRh$_2$O$_4$, a combined experimental and theoretical work showed that the S = 3/2 spins are unfrustrated and display static and dynamic properties \cite{20}. Tetragonally distorted CuRh$_2$O$_4$ shows an incommensurate magnetic order for the S = 1/2 spins and the presence of sizable quantum effects \cite{20}. A spin-orbit entangled paramagnetic state is suggested in NiRh$_2$O$_4$ \cite{21, 30, 31}. Studies on magnetically diluted Cu$_{1-x}$Zn$_x$Rh$_2$O$_4$ highlight spin transition triggered by an enhancement of preceding spin fluctuations \cite{32}. At the same time, it shows the suppression of orbital order on the octahedral sites through the percolative manner. In general, geometrically frustrated structures with 4d/5d elements (SOC) are ideal for realizing exotic phenomena. Therefore, exploring new frustrated structures (geometrical frustration and competing magnetic interactions) coupled with SOC is essential.

In this article, we present the detailed structural (using X-ray powder diffraction, neutron powder diffraction, and electron microscopy), magnetic (magnetization), thermal (specific heat), and magneto-dielectric studies on a mixed 3d - 4d spinel oxide with composition CoFeRhO$_4$. In this material, non-magnetic Rh$^{3+}$ cations occupy octahedral B sites; however, they can be a source of SOC in the structure. An insulating and ferrimagnetic ground state is observed near room temperature (T$_C$ = 355 K). Detailed magnetic and magneto-dielectric measurements highlight the room-temperature exchange bias effect and the magnetodielectric effect in this material.

\section{Experimental Details}
\textbf{Synthesis.} We have used the standard solid-state reaction route to synthesize the polycrystalline CoFeRhO$_4$ materials. Stoichiometric amounts of Co$_3$O$_4$ (99.9\%) Fe$_2$O$_3$ (99.999\%) and Rh (99.9\%) metal powders were used for the synthesis. The stoichiometric amounts of the starting materials were mixed with a mortar pestle. The materials were heated several times and the final sintering was performed at 1473 K for 36 h.

\textbf{X-ray Powder Diffraction and Neutron Diffraction.} X-ray diffraction (XRD) using the powder form of the material was collected at ambient temperature (RT) using a PANalytical diffractometer (Cu-K$_{\alpha}$, $\lambda$ = 1.54056 \text{\AA}). Neutron diffraction (NPD) data for the powdered CoFeRhO$_4$ sample were collected at various temperatures (12 K, 100 K and 250 K) in JEEP II, Kjeller Reactor, Norway ($\lambda$ = 1.5538 \text{\AA}). We have performed the Rietveld refinements of the diffraction patterns (XRD and NPD) using the fullProf suite software. 

\textbf{Electron Microscopy.} We have performed electron microscopy measurements (high-angle annular dark-field scanning transmission electron microscopy (HAADF-STEM) and electron diffraction) to explore the crystal structure of CoFeRhO$_4$. HAADF-STEM and electron diffraction were carried out using a JEOL ARM-200F cold FEG probe image aberration corrected 200 kV microscope. The instrument is also equipped with a solid-angle CENTURIO EDX detector. 

\textbf{Magnetism and Specific Heat.} Direct current magnetization (temperature and magnetic field dependent) measurements were done in a Quantum Design superconducting quantum interference device (MPMS 3). Magnetic measurements were conducted in field-cooled (FC) and zero-field-cooled (ZFC) modes. Specific heat was measured (2 K - 400 K) in a physical property measurement system (PPMS, Quantum Design) without a magnetic field.

\textbf{Dielectric Measurements.} The temperature and magnetic field-dependent dielectric measurements were performed employing an LCR meter (Agilent 4284A) and a sample insert for a PPMS (Quantum Design). 

\section{Results and Discussion}
\begin{figure}
\includegraphics[width=0.90\columnwidth]{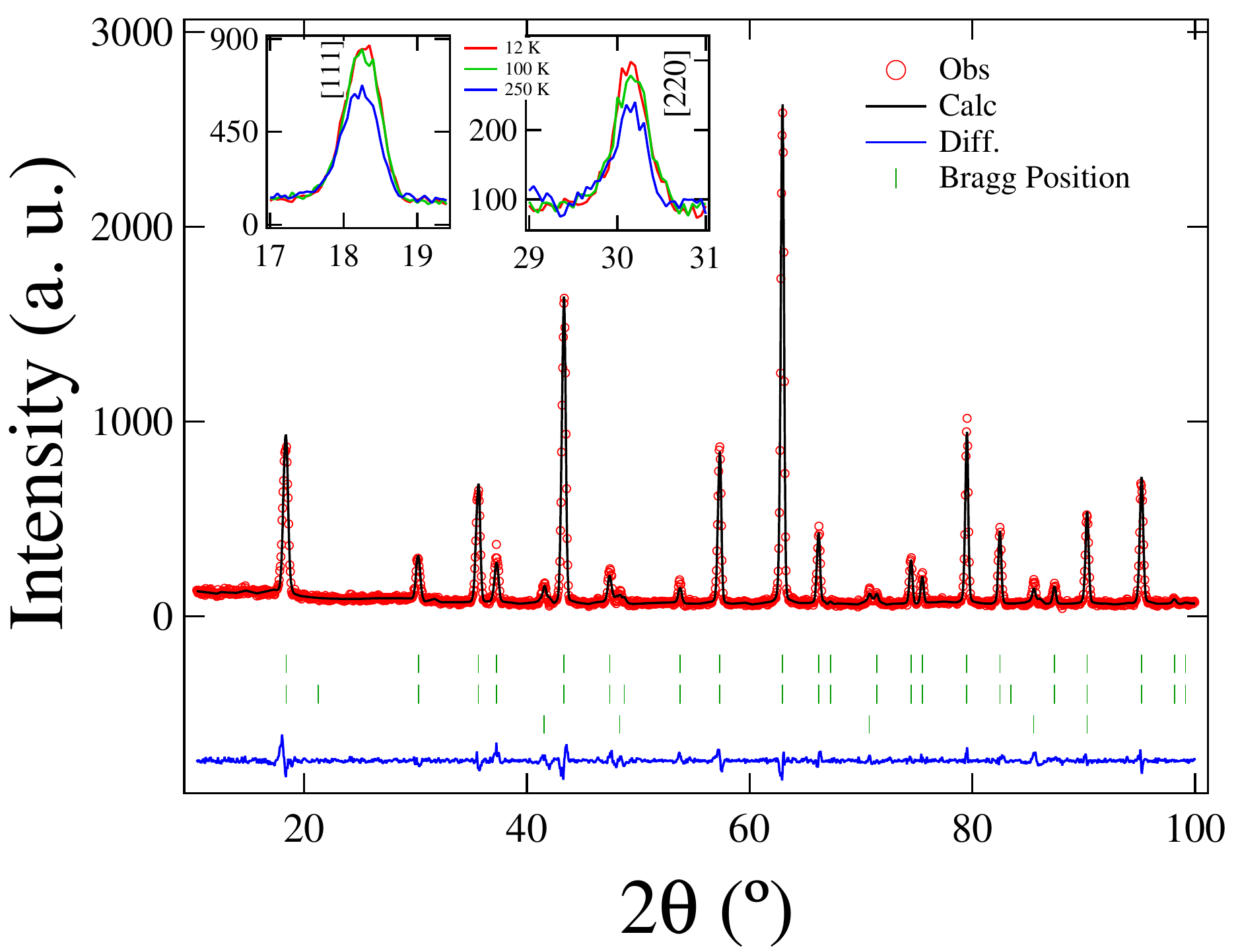}
\caption{\label{Figure 2:NPD}Rietveld refinement plot of the Neutron powder diffraction (NPD) pattern collected at 12 K for CoFeRhO$_4$. The lower ticks highlight the magnetic peak positions (k = 0, 0, 0). Insets in the figure show the enhancement of the [111] and [220] NPD peaks with a lowering of the temperature.}
\end{figure}

\begin{figure}
\includegraphics[width=0.88\columnwidth]{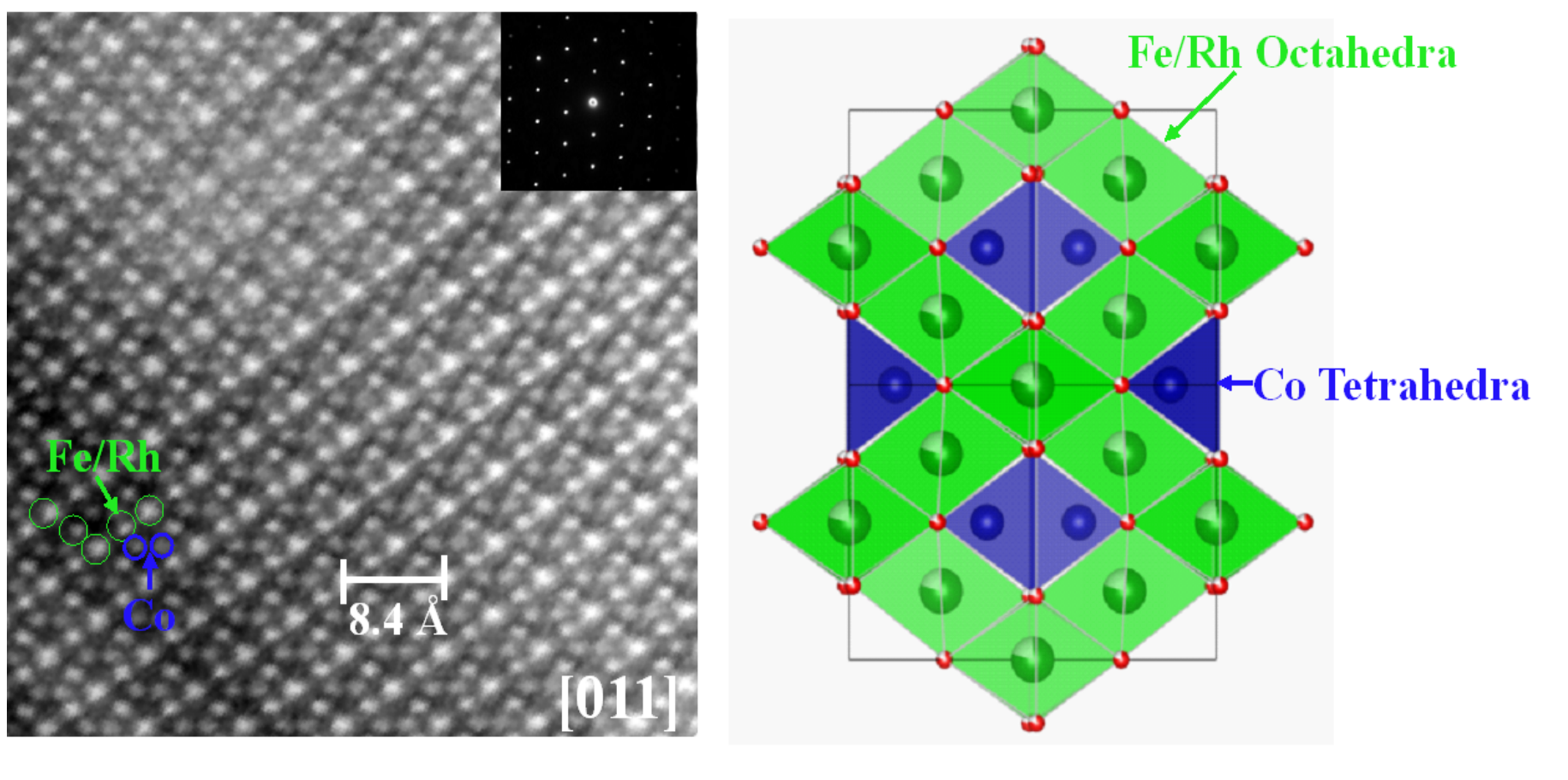}
\caption{\label{Figure 3:STEM}HAADF image and crystal structure along the [011] zone axis for CoFeRhO$_4$, collected at room temperature. Fe, Rh, and Co cations are highlighted in the figure. The corresponding electron diffraction pattern is shown in the inset.}
\end{figure}

Preliminary Rietveld refinement uses the room-temperature XRD pattern of CoFeRhO$_4$. It shows that the sample crystallizes in a cubic structure with a space group $Fd\overline{3}m$, which is isostructural with CoFe$_2$O$_4$ \cite{33, 34}. However, a small amount of Rh is detected (main peak at 2$\theta$ $\approx$ 41 degrees, Fig. S1 in the supporting information (SI)) \cite{SI}  in the RT XRD pattern for CoFeRhO$_4$. Further, to explore the detailed nuclear and magnetic structure, we have collected the NPD patterns at 12 K, 100 K, and 250 K. Therefore, we will use the results of the NPD Rietveld refinements to describe the crystal structure of CoFeRhO$_4$. Similar to the XRD pattern, a small amount of Rh is also detected in the NPD patterns. Fig.\ref{Figure 2:NPD} illustrates the plot of the NPD Rietveld refinement for the pattern collected at 12 K. The plots of the NPD Rietveld refinements for the patterns collected at different temperatures are provided in the supporting information (Figure S2 in SI) \cite{SI}. The structural parameters obtained from the 12 K - NPD refinement for CoFeRhO$_4$ are summarized in Table 1. The crystal structure for CoFeRhO$_4$ is highlighted in Fig.\ref{Figure 1:SRT}. Rh atoms and most of the Fe cations occupy the octahedrally coordinated B site (16d (0.5, 0.5, 0.5)) of the structure. Co-cations occupy the eightfold tetrahedral A sites (8a (0.125, 0.125, 0.125)). Oxygen atoms occupy the 32e (x, x, x) Wycoff positions. The refinement of the oxygen occupancy indicates that there is no deviation from the full occupancy. Rh shows a small deviation from full occupancy (Occupancy = 0.98 (2)). Fig.\ref{Figure 3:STEM} shows the STEM- HAADF image along the [011] direction for CoFeRhO$_4$. The observed STEM images agree with the expected crystal structure of the spinel material. The lattice parameter obtained from STEM images (8.4 \AA) matched the values found from the NPD (Table 1) and RT-XRD refinements.

\begin{figure}
\includegraphics[width=0.85\columnwidth]{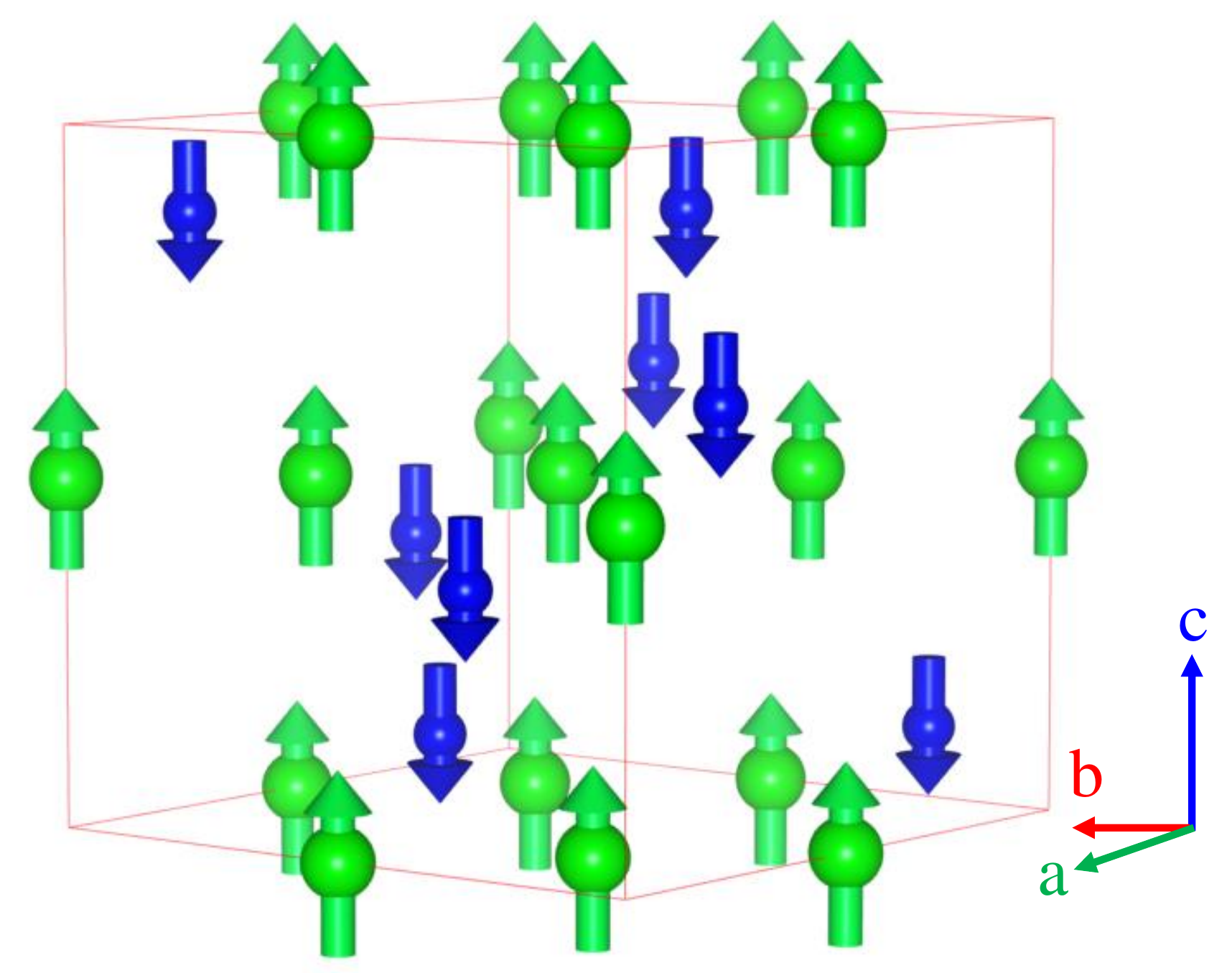}
\caption{\label{Figure 4:Mag} The observed magnetic structure (k = 0, 0, 0) for CoFeRhO$_4$. A collinear ferrimagnetic ordering scheme is observed in the NPD refinements.}
\end{figure}

The structural (Rietveld) refinement of the 250 K NPD highlights the existence of magnetic peaks at that temperature. These magnetic ordering-related peaks show an increase in intensity with decreasing temperature (inset in Fig.\ref{Figure 2:NPD}). The position of all magnetic peaks for CoFeRhO$_4$ coincides with the allowed nuclear reflections. Therefore, we describe the magnetic structure using the propagation vector k = (0,0,0). For the long-range magnetic order, the best fit to the NPD magnetic reflections is achieved using a collinear ferrimagnetic ordering model (Fig. \ref{Figure 4:Mag}). A similar magnetic ordering scheme was proposed for the CoFe$_2$O$_4$ material \cite{33}. The absence of a (200) magnetic peak discards the non-collinear arrangement of the spins. In general, any long-range spin canting would result in the appearance of the (200) magnetic Bragg peak. We do not observe the (200) peak at 21.27 $\degree$ ((400) peak is observed at $2\theta$ = 43.43 $\degree$). 

\begin{table}[h!]
\caption{Structural parameters for Ba$_{2}$ScRuO$_{6}$ at RT extracted from Rietveld refinement of powder XRD diffraction data}
\label{Table1}
 Space group~~$Fd\overline{3}m$, \textit{a = b = c} = 8.4207 (1) \AA\\
R$_P$ = 6.98, R$_{WP}$ = 9.01, $\chi$$^2$ = 1.03,R$_{Bragg}$ = 2.79\\
R$_{Mag}$ = 5.29, M (T$_d$) = 3.4 (1) $\mu$$_B$, M (Oct.) = 4.1 (1) $\mu$$_B$
\begin{center}
\small\addtolength{\tabcolsep}{-1pt}
\begin{tabular}{c c c c c c c} \hline
Atom & Wyck. Pos. & x & y & z & Occu. & B$_{iso}$\\
\hline
\\[0.5ex]                                  
Co(T$_d$)  & 8a & 0.125 & 0.125 & 0.125 & 0.88 (2) & 0.1 (1)\\   
Fe(T$_d$)  & 8a & 0.125 & 0.125 & 0.125 & 0.12(2) & 0.1(1)\\
Co(Oct.)  & 16d & 0.5 & 0.5 & 0.5 & 0.065(5) & 0.11(3)\\   
Fe(Oct.) & 16d  & 0.5 & 0.5 & 0.5 & 0.44(1) & 0.11(3)\\
Rh(Oct.) & 16d & 0.5 & 0.5 & 0.5 & 0.49(1) & 0.11(3)\\
O          & 32e & 0.249(1) & 0.249(1) & 0.249(1) & 1 & 0.10(3)\\

\\[0.5ex]
\hline\\
\end{tabular}
\par\medskip\footnotesize
\end{center}
\end{table}

Magnetic moments in different sublattices (M (T$_d$) - tetrahedral A sites, M (Oct) - octahedral B sites) obtained from refinements are listed in Table 1. The resultant magnetic moment per formula unit obtained from the NPD refinements is 0.6 (1) $\mu$$_B$ at T = 12 K. However, a discrepancy is observed between the magnetic moments' experimental and theoretical values in the tetrahedral and octahedral sites. Theoretically calculated magnetic moments in different sublattices are M$_A$ = $0.88\times3.5 + 0.12 \times 5$ = 3.66 $\mu$$_B$ and M$_B$ = $0.87 \times 5 + 0.13 \times 3.5)$ = 4.81 $\mu$$_B$, calculated with M$_F$$_e$$^3$$^+$ = 5 $\mu$$_B$ and M$_C$$_o$$^2$$^+$ = 3.5 $\mu$$_B$. Reduced magnetic moments could result from local disorder (and/or local canting) of spins. The complex cationic distribution and the existence of nonmagnetic Rh in the structure can originate competing magnetic interactions and can create nonuniform spin canting. A similar reduced magnetic moment due to the local spin canting is also observed for the CoFe$_2$O$_4$ and Ti-doped CoFe$_2$O$_4$ materials \cite{33,35}. However, a collinear ferrimagnetic long-range magnetic ordering scheme is suggested for both of these materials.\\

\begin{figure}
\includegraphics[width=0.85\columnwidth]{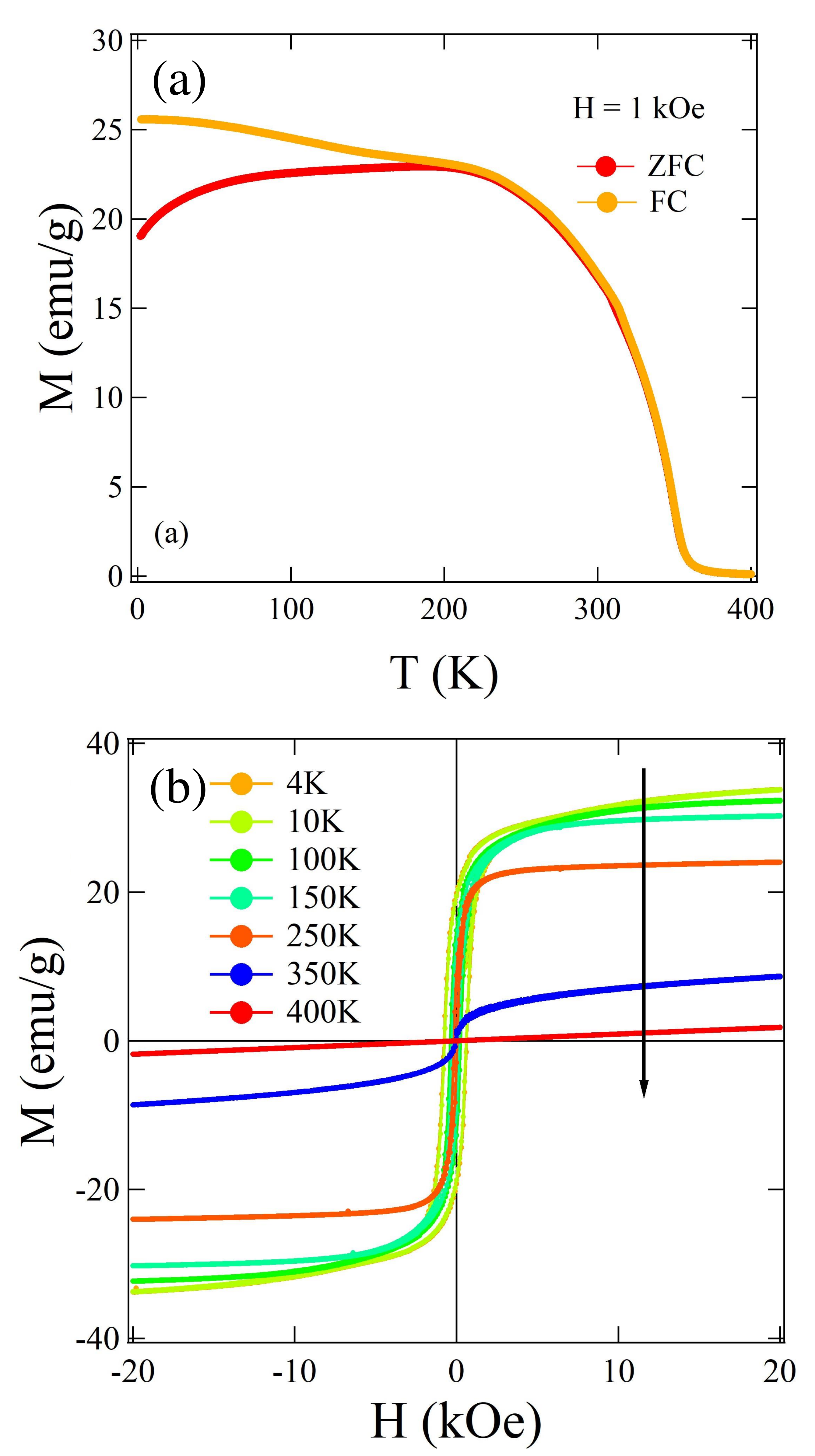}
\caption{\label{Figure 5:MT} (a) Temperature-dependent magnetization (M - T) and (b) magnetic field variation of the magnetization (M - H) for CoFeRhO$_4$. Ferrimagnetic behavior is observed for CoFeRhO$_4$.}
\end{figure}

The FC and ZFC magnetic moment (M-T) temperature variation measured at 0.1 T for CoFeRhO$_4$ is presented in \ref{Figure 5:MT}. The M-T shows a sharp upturn at 355 K, highlighting the onset of the ferrimagnetic transition at that temperature. Upon further lowering the temperature bifurcation between FC and ZFC magnetization (M$_{FC}$ and M$_{ZFC}$) is visible at 250 K. Below 250 K, M$_F$$_C$ shows an increasing trend with decreasing temperature. Several competing magnetic interactions exist in CoFeRhO$_4$. The complex crystal structure (diamond and pyrochlore lattice) and the distribution of magnetic and nonmagnetic cations in two different sublattices can create competing magnetic interactions and frustration. For example, A-site magnetic cations are surrounded by 12 B-site magnetic and non-magnetic first neighbours, which can be Fe$^{3+}$ or Co$^{2+}$ or Rh$^{3+}$; any B-cation is surrounded by 6 A and 6 B-sited cations. Therefore, the bifurcation and the rise in M$_{FC}$ at a lower temperature could be due to the frustration in the spinel structure. However, the ferrimagnetic transition in CoFeRhO$_4$ is much lower than in CoFe$_2$O$_4$. In CoFe$_2$O$_4$, the collinear ferrimagnetic transition is observed at 800 K with a strong intersublattice AFM superexchange interaction (J$_{AB}$ = - 12.39 k$_B$) \cite{33}. Introducing non-magnetic Rh into the structure could dilute the magnetic interactions by decreasing the strength of inter- and intra-sublattice superexchange interactions and, therefore, can reduce the magnetic ordering temperature. Figure 5(b) shows the magnetization as a function of the magnetic field (M - H) at various temperatures measured in the ZFC mode. Above the magnetic transition temperature (at 400 K), a linear paramagnetic type M-H behavior is observed. However, at 350 K, the M-H loop illustrates non-linearity at low magnetic fields, indicating the appearance of the magnetically ordered state. As we further lower the temperature, saturated ferrimagnetic-type M - H loops are observed. The saturation magnetization and coercive field (H$_C$) enhance with decreasing the temperature.

Figure 6 highlights the temperature variation of total specific heat for CoFeRhO$_4$. Magnetic ordering usually involves an entropy change, resulting in specific heat anomalies. The C$_P$ vs T plot shows an anomaly at 355 K, indicating a true phase transition. This also confirms the magnetic ordering temperature for CoFeRhO$_4$. The low-temperature part of the specific heat (2-20 K) can be well represented by eqn. \ref{eqn1:SP}:

\begin{equation}
C_{P} = \gamma T + \beta T^{3}  ,
\label{eqn1:SP}
\end{equation}

where $\gamma$ = Sommerfeld coefficient and $\beta$ = lattice contributions to the specific heat. The obtained $\gamma$ =  0.015 Jmol$^{-1}$K$^{-2}$. However, the total specific heat at higher temperatures contains both the phonon and magnetic parts ($C_{P} = C_{ph} + C_{m}$). A combined Einstein-Debye model can generally estimate the total phonon contribution in the specific heat. Therefore, to estimate the magnetic contributions ($C_{m}$) in the specific heat, we fit the data with the Einstein-Debye model \cite{36}: 
\begin{equation}
C_{ph} = 9Rx_D^{-3}\int_{0}^{x_D}\frac{x^4 e^x}{({e^{x}-1})^2}  dx + R\sum_{i=1}^{2}a_{i}\frac{x_{Ei}^2 e_{Ei}^x}{({e_{Ei}^{x}-1})^2}
\label{eqn2:ED}
\end{equation}

\begin{figure}
\includegraphics[width=0.88\columnwidth]{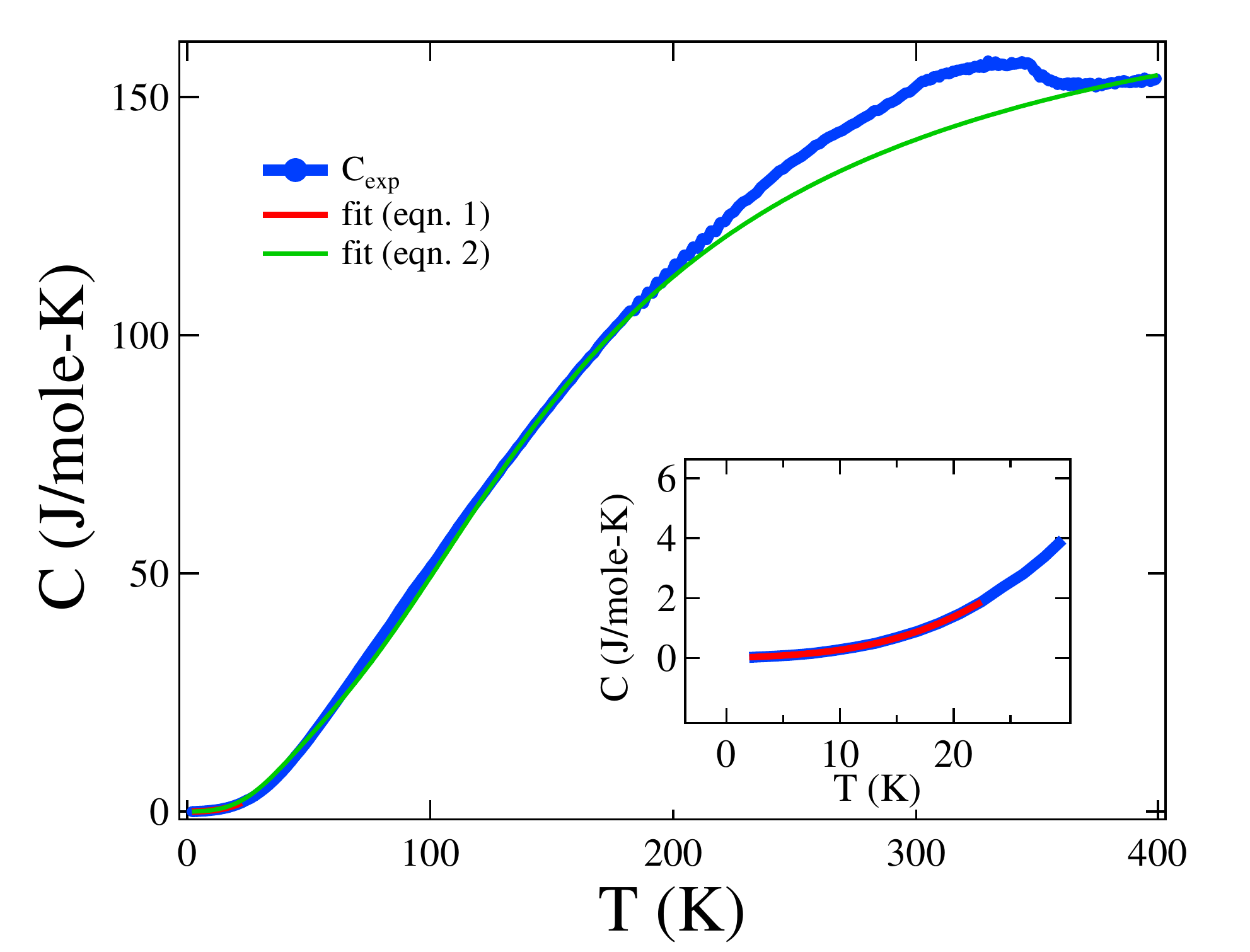}
\caption{\label{Figure 6:SP} Temperature variation of the specific heat (C) for CoFeRhO$_4$. An anomaly due to the magnetic ordering is observed at 355 K. The green line highlights the fitted curve using Eqn.\ref{eqn2:ED}. Low-temperature C is also fitted using eqn.\ref{eqn1:SP} and shown in the inset of the figure.}
\end{figure}

Where x$_D$ = $\theta$$_D$/T and x$_{Ei}$ = $\theta$$_{Ei}$/T and, Debye temperature = $\theta$$_D$ and $\theta$$_Ei$ = Einstein temperature. R = 8.31 J/mol K, and a$_i$ = degree of freedom for each Einstein mode. The fitted curve is shown in figure 6. The best fit is obtained with $\theta$$_D$ = 632 K, $\theta$$_{E1}$ = 177 K and $\theta$$_{E2}$ = 740 K. Then the contribution related to the magnetic ordering (C$_m$) in the specific heat is calculated by subtracting C$_ph$ from the measured total specific heat data. Figure 7 highlights the C$_m$ - T plot, showing a sharp maximum at T = 355 K. The entropy change due to the magnetic ordering can be calculated by using the following formula:
\begin{equation}
S_{m} (T) = \int_{0}^{T}\frac{C_m (T)}{T} dT
\label{eqn3:MSP}
\end{equation}

The $S_{m} (T)$ vs T plot is shown in figure 7.  Magnetic entropy increases with increasing temperature and shows a saturation value of $\approx$ 4.4 J mole$^{-1}$ K$^{-1}$ (above 355 K), and this is less than the theoretically estimated magnetic entropy for CoFeRhO$_4$ ($S_{m} (T)$ = R $\ln{(2S+1)}$ = 26.41 J mole$^{-1}$ K$^{-1}$, Co$^{2+}$ adopts the e$_g$$^{4}$t$_2g$$^{3}$ electronic configuration with S = 3/2,  Fe$^{3+}$ adopts the e$_g$$^{2}$t$_2g$$^{3}$ electronic configuration with S = 5/2). In general, the frustration of the magnetic cations near and above the magnetic transition temperature can reduce the entropy contribution to the magnetic ordering. In CoFeRhO$_4$, geometrical frustration in the structure (diamond and pyrochlore lattice) and the distribution of magnetic and non-magnetic cations in two different sublattices could cause frustration of magnetic cations. The reduced magnetic moments in the neutron powder diffraction experiments point to spins' local disorder (and/or local canting). Therefore, the reduced entropy related to magnetic ordering can be attributed to the frustration of the magnetic cations \cite{37}. 
\begin{figure}
\includegraphics[width=0.90\columnwidth]{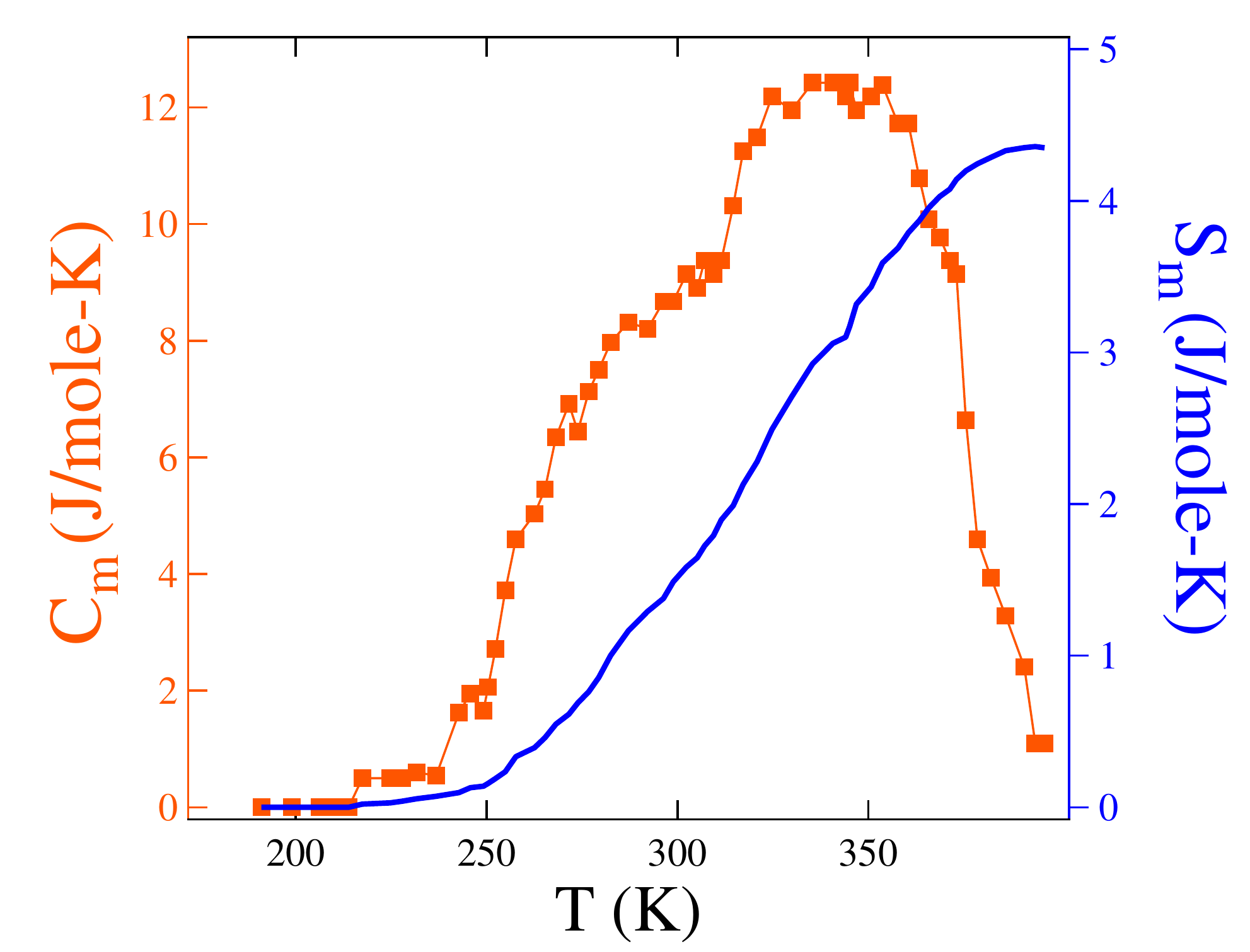}
\caption{\label{Figure 7:SPMag} Temperature variation of the magnetic component of the specific heat (C) (left axis) and calculated magnetic entropy (right axis) for CoFeRhO$_4$}
\end{figure}

Oxides geometrical frustration and ferrimagnetic transition often result in exciting magnetoelectric/magnetodielectric coupling. Geometric frustration (local spin canting) in a magnetic system is proven to be a key ingredient for magneto(di)electric coupling. For example, the triangular Ising lattice Ca$_3$Co$_2$O$_6$ shows magneto dielectric coupling below the ferrimagnetic ordering temperature (24 K) \cite{38}. Haldane spin-chain system Dy$_2$BaNiO$_5$ shows a magneto-(di)electric effect below the long-range ordering temperature (58 K) \cite{39}. In spinels, the magnetodielectric effect is observed from the beginning of the ferrimagnetic ordering in CoCr$_2$O$_4$ (T$_C$ = 96 K) \cite{40, 41} and NiCr$_2$O$_4$ \cite{42}. MnCr$_2$O$_4$ shows magneto dielectric effect below the ferrimagnetic ordering temperature (43 K) \cite{43}. However, the magneto-(di)electric effect in ferrimagnetic materials near room temperature is rare. The near-room-temperature magnetodielectric effect is important due to its possible applications in spintronics devices, magnetically accessible ferroelectric random-access memories, and communication technology. Figure 8 shows the temperature variation of the real part of the dielectric constant ($\epsilon _r$) for CoFeRhO$_4$. The dielectric loss (tan$\delta$) is highlighted in the inset of Figure 8. The extremely low values of dielectric loss (tan$\delta$) highlight the insulating behaviour of CoFeRhO$_4$. It also excludes the possibility of extrinsic Maxwell-Wagner-like behaviour appearing because of the leakage currents. However, the frequency dispersion of $\epsilon _r$ and tan$\delta$ started above 320 K, indicating the growing Maxwell-Wagner-type relaxation or other sources of conductivity around 320 K. Nonetheless, the dielectric constant is intrinsic below 320 K. In addition to a small hump near the magnetic ordering temperature (335 K), two anomalies at 220 K and 50 K are observed in the $\epsilon _r$ - T plots.  In particular, the magnetisation measurements also observed similar anomalies at 220 K and 50 K. 
\begin{figure}
\includegraphics[width=0.88\columnwidth]{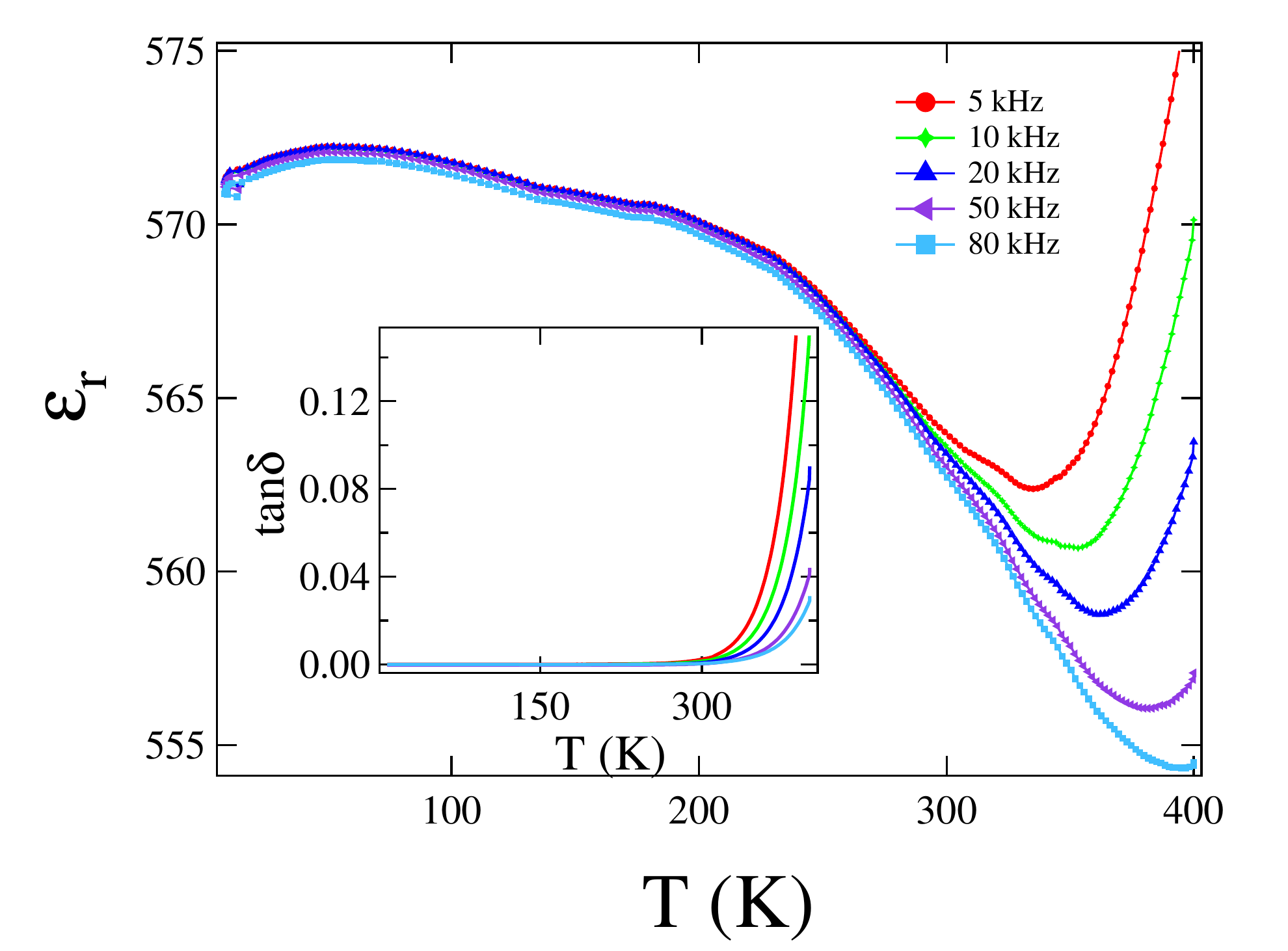}
\caption{\label{Figure 8:DE} Temperature variation of the real part of the dielectric constant ($\epsilon _r$) at several fixed frequencies for CoFeRhO$_4$. The inset in the figure shows the dependence of the dielectric loss as a temperature function.}
\end{figure}

As observed from our nuclear and magnetic structure refinements, heat capacity, and magnetization studies, CoFeRhO$_4$ hosts an exciting combination of complex crystal structure (diamond and pyrochlore lattice) and competing magnetic interactions. The reduced magnetic moments observed in the NPD refinements highlight the possibility of local spin canting. Therefore, the geometrical frustration, competing magnetic interactions, and the local spin canting can originate magnetic frustration in the structure.  The magnetic frustration could be why these anomalies are observed below the long-range magnetic ordering. To understand the effect of the magnetic field on the dielectric constant, we have performed magnetic field dependence measurements of the dielectric constant. Figure 9 shows the temperature variation of $\epsilon _r$ measured in the presence of different magnetic fields.
Interestingly, the dielectric constant increases in the presence of a magnetic field below the magnetic ordering temperature, and a positive magnetodielectrictance effect is observed. The starting of the magneto(di)electric effect below the long-range ferrimagnetic ordering temperature indicates that the observed magneto(di)electric effect is related to the magnetic ordering of CoFeRhO$_4$. The change in the dielectric data with the magnetic field is most pronounced near 220 K. The magneto-dielectric constant ($\Delta \epsilon _r = [\epsilon _r (H) - \epsilon _r (H = 0)]/ \epsilon _r (H = 0)$) and its dependence on the magnetic field are shown in the inset of figure 9. As discussed, the magnetic frustration in the structure (geometric frustration, competing magnetic interactions, and local spin canting) could be the origin of the increase in $\epsilon _r$ with the magnetic field at that temperature.

\begin{figure}
\includegraphics[width=0.85\columnwidth]{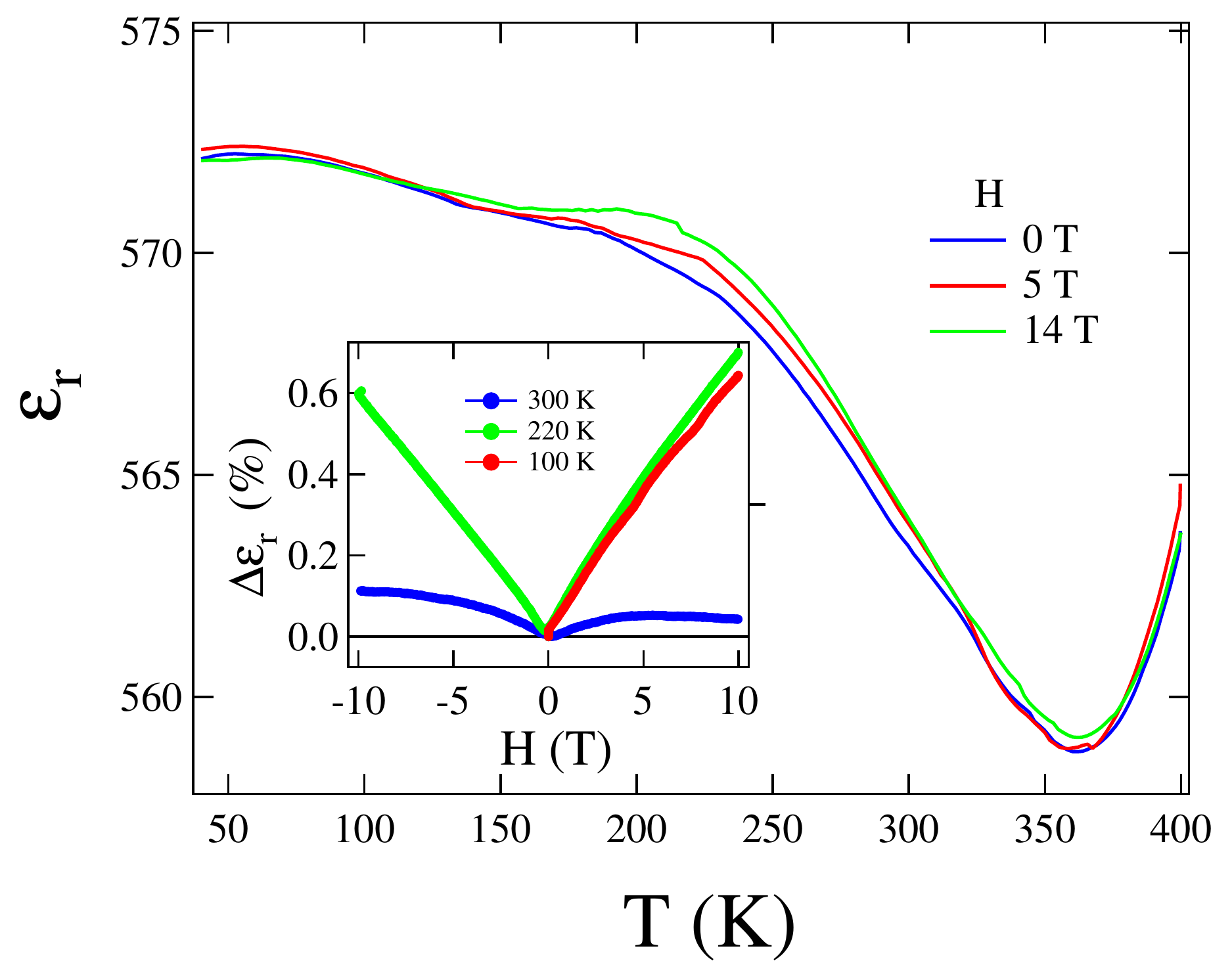}
\caption{\label{Figure 9:MagDE} Temperature dependent dielectric constant (real part, $\epsilon _r$) at different magnetic fields (H = 0 T, 5 T, and 14 T) for CoFeRhO$_4$. The inset highlights the magnetic field variation of the dielectric constant ($\Delta \epsilon _r = [\epsilon _r (H) - \epsilon _r (H = 0)]/ \epsilon _r (H = 0)$) at different temperatures.}
\end{figure}

In recent times, complex magnetic materials (for instance, ferrimagnetic systems Mn$_{3-x}$Pt$_x$Ga \cite{44}, Ba$_2$Fe$_{1.12}$Os$_{0.88}$O$_6$ \cite{45} and SrFe$_{0.15}$Co$_{0.85}$O$_{2.62}$) \cite{46}, in particular, those having a magnetic transition temperature near room temperature have attracted significant interest in exploring the exchange bias effect (EB). The EB effect shifts the isothermal magnetization loop, becoming asymmetric and shifting along the field axis. This effect has important technological applications, such as in the development of magnetic sensors, magnetic recording read heads \cite{47}, random access memories \cite{48}, and other spintronic devices \cite{49,50}. Figure 10 highlights the magnetic field dependence of magnetization at 100 K, measured in FC mode (cooling field, $H_{FC}$ = 3 T). It shows a clear shift towards the left field (negative magnetic field). The EB field (H$_{EB}$) can be calculated from the shift of the hysteresis loop using the following equation
\begin{equation}
H_{EB} = -(H_{C(L)} + H_{C(R)})/2
\label{eqn4:EB}
\end{equation}

 $(H_{C(L)}$ and $H_{C(R)}$ are the intercepts with the positive (right) and negative (left) field axis. The calculated exchange bias field H$_{EB}$ = 65 Oe at 100 K. The exchange bias field is almost constant with varying cooling fields ($H_{FC}$, ranging from 1 T to 7 T). To investigate the temperature evolution of the exchange bias effect and to explore whether the exchange bias property in CoFeRhO$_4$ is correlated with magnetic ordering, we have measured the temperature variation of the EB field. The material was cooled to the measuring temperatures in a magnetic field of 3 T. Then the M-H loops were measured between $\pm$0.5 T. Fig. \ref{Figure 11:EBT} highlights the temperature evolution of the EB field for CoFeRhO$_4$. Interestingly, the EB effect emerges just below the long-range ferrimagnetic ordering (350 K); however, with decreasing temperature, H$_{EB}$ shows almost constant behavior down to 10 K. In ferrimagnetic systems, the presence of uncompensated magnetic moments in a compensated AFM host is proven to be very effective in designing the exchange bias effect (for instance, Mn$_{3-x}$Pt$_x$Ga \cite{44}, Ba$_2$Fe$_{1.12}$Os$_{0.88}$O$_6$) \cite{45}. 
 
\begin{figure}
\includegraphics[width=0.86\columnwidth]{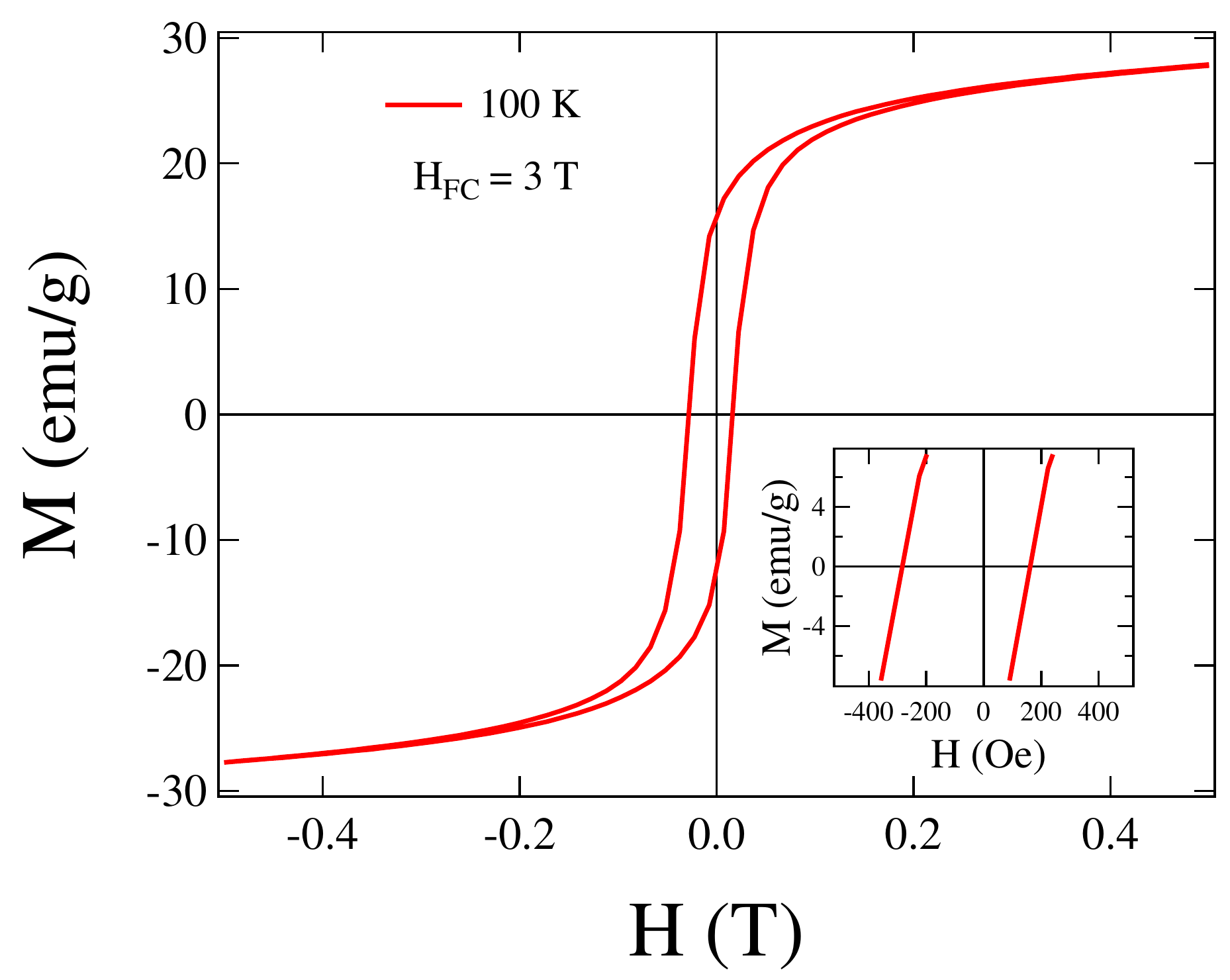}
\caption{\label{Figure 10:EB} $M$-$H$ loop of CoFeRhO$_4$ at 100 K measured in field cooled (FC) mode ($H_{FC}$ = 3 T). The inset highlights the enlarged central part of the FC $M$-$H$ loop. A shift along the left field axis is highlighted here.}
\end{figure}

 In this spinel structure, there is an uncompensated magnetic moment due to the distribution of two different magnetic cations (Co$^{2+}$ and Fe$^{3+}$) in two different sublattices. In CoFeRhO$_4$, as observed from the neutron diffraction refinements (table 1), below the magnetic ordering temperature, the magnetic moment of B sublattice (M (Oct)) dominates over the M (T$_d$). Under FC conditions, the magnetic field orients the net magnetic moment of the sublattices along the field direction. Now, as we gradually reverse the direction of the magnetic field, the magnetic moments of the B sublattice (M (Oct)) do not easily reverse its direction as they are antiferromagnetically coupled with the A-sublattice (M (T$_d$)). This pinning effect of irreversible uncompensated spins (Octahedral B sublattice) develops the unidirectional anisotropy, and therefore, exchange bias is observed in CoFeRhO$_4$.

\section{Conclusion}
To summarize, we synthesized and examined the detailed structural, magnetic, thermal (specific heat), magneto-dielectric, and magnetic exchange bias properties of CoFeRhO$_4$, a mixed 3d-4d spinel oxide. We used various diffraction techniques, such as RT-XRD, NPD, RT-electron diffraction, and STEM, to explore the material's structural details. The material crystallizes in a cubic structure with a space group of $Fd\overline{3}m$. Our measurements show a long-range ferrimagnetic ordering with an onset at 355 K, which is explained by a collinear ferrimagnetic ordering model with k = [0, 0, 0]. However, we observe reduced magnetic moments in the refinement, possibly due to the spins, local disorder, local canting, or frustration. Magnetic frustration can also reduce magnetic entropy, as reflected in the specific heat measurement. The magnetic entropy increases with temperature and reaches a saturation value of 4.4 J mole$^{-1}$K$^{-1}$ (above 355 K), but it is less than the theoretically estimated value.
Furthermore, our dielectric and FC magnetization measurements show the appearance of two technologically significant phenomena near room temperature: the magnetodielectric effect and the exchange bias effect. These effects appear below the magnetic ordering temperature (355 K) and can originate from magnetic frustration, collinear ferrimagnetic ordering, and uncompensated magnetic moments. The uncompensated magnetic moments create unidirectional anisotropy, resulting in an exchange bias effect. Importantly, the presence of room temperature ferrimagnetism, EB, and the magneto(di)electric effect demonstrates the potential of CoFeRhO$_4$ in developing materials for sensors or spintronics applications operating at room temperature.

\begin{figure}
\includegraphics[width=0.92\columnwidth]{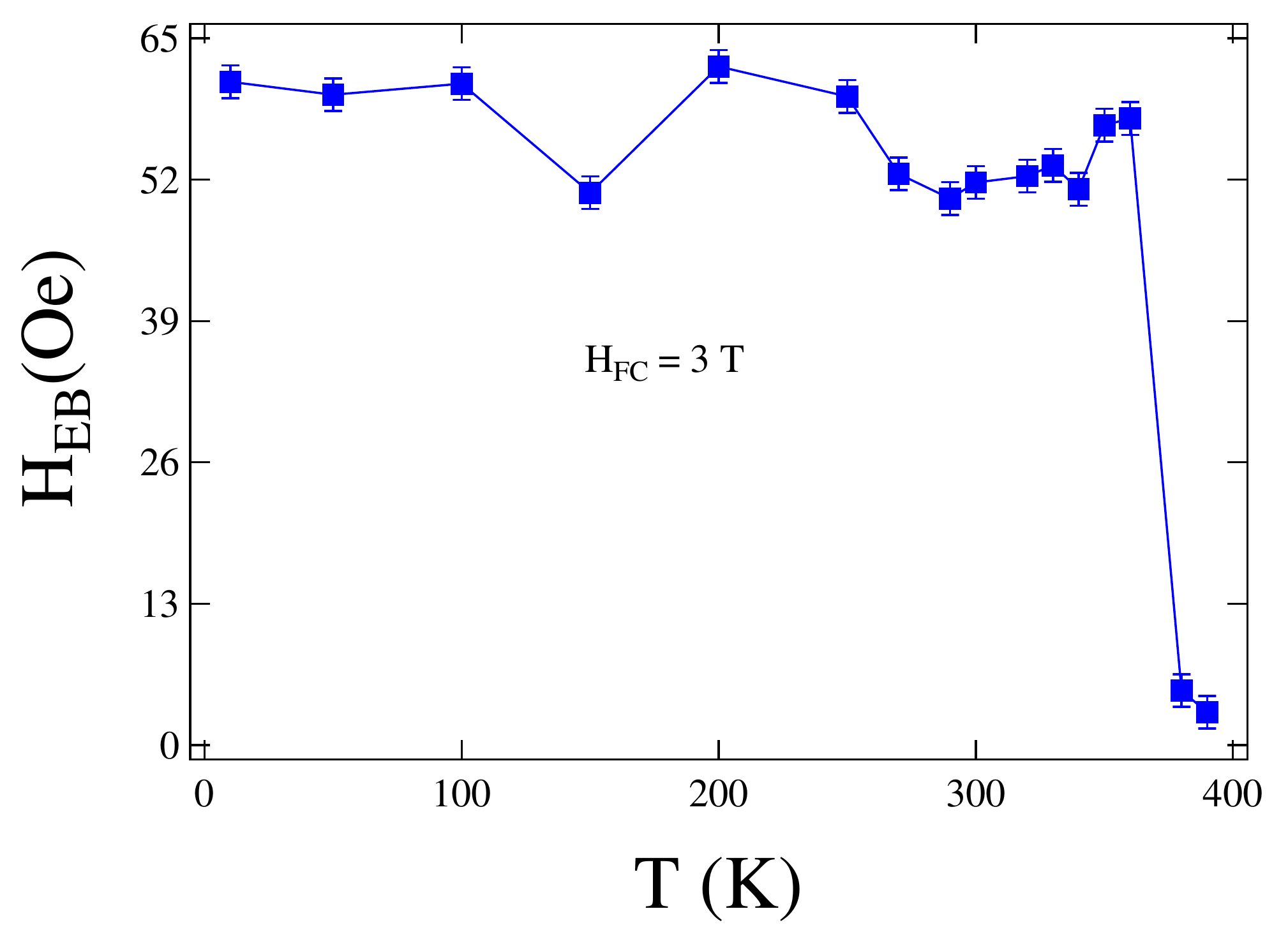}
\caption{\label{Figure 11:EBT} Temperature variation of the exchange bias field ($H_{EB}$) for CoFeRhO$_4$. The exchange bias effect started to appear below the long-range ferrimagnetic ordering (350 K).}
\end{figure}

\section{Acknowledgments}

R. P. S. acknowledges the Science and Engineering Research Board (SERB), Government of India, for the CRG/2019/001028 Core Research Grant. S. M. acknowledges the SERB, Government of India, for the SRG/2021/001993 Start-up Research Grant.

\end{document}